\documentclass[12pt,preprint,dvips]{aastex}

\shorttitle{H$\alpha$ luminosity function of Abell 521}
\shortauthors{Umeda et al.}

\begin{document}


\title{THE H$\alpha$ LUMINOSITY FUNCTION OF THE GALAXY CLUSTER ABELL 521
AT $z$ = 0.25\altaffilmark{1}}

\author{Kazuyoshi Umeda\altaffilmark{2},
        Masafumi Yagi\altaffilmark{3},
        Sanae F. Yamada\altaffilmark{2},
        Yoshiaki Taniguchi\altaffilmark{2}, \\
        Yasuhiro Shioya\altaffilmark{2},
        Takashi Murayama\altaffilmark{2},
        Tohru Nagao\altaffilmark{2},
        Masaru Ajiki\altaffilmark{2},
        Shinobu S. Fujita\altaffilmark{2}, \\
        Yutaka Komiyama\altaffilmark{4},
        Sadanori Okamura\altaffilmark{5,6}, and
        Kazuhiro Shimasaku\altaffilmark{5,6}}

\altaffiltext{1}{Based on data collected at the Subaru Telescope, which
is operated by the National Astronomical Observatory of Japan.}

\altaffiltext{2}{Astronomical Institute, Graduate School of Science,
Tohoku University, Aramaki, Aoba, Sendai 980-8578, Japan}

\altaffiltext{3}{National Astronomical Observatory of Japan, Mitaka, Tokyo
181-8588, Japan}

\altaffiltext{4}{Subaru Telescope, National Astronomical Observatory of
Japan, 650 N. A'ohoku Place, Hilo, HI 96720}

\altaffiltext{5}{Department of Astronomy, Graduate School of Science,
University of Tokyo, Tokyo 113-0033, Japan}

\altaffiltext{6}{Research Center for the Early Universe, School of
Science, University of Tokyo, Tokyo 113-0033, Japan}

\begin{abstract}

We present an optical multicolor-imaging study of the galaxy cluster
 Abell 521 at $z = 0.25$, using Suprime-Cam on the Subaru Telescope,
 covering an area of $32 \times 20$ arcmin$^2$ ($9.4 \times 5.8 h_{50}^{-2}$
 Mpc$^2$ at $z = 0.25$).
Our imaging data taken with both a narrow-band filter, $NB816$
 ($\lambda_0 = 8150$\AA\ and $\Delta \lambda = 120$\AA), and broad-band
 filters, $B,V,R_{\rm C}, i^\prime$, and $z^\prime$ allow us to find 165
 H$\alpha$ emitters.
We obtain the H$\alpha$ luminosity function (LF) for the cluster
galaxies within 2 Mpc; the Schechter parameters are $\alpha = -0.75 
\pm 0.23$, $\phi^\star = 10^{-0.25 \pm 0.20}$ Mpc$^{-3}$, and
$L^\star = 10^{42.03 \pm 0.17}$ erg s$^{-1}$.
Although the faint end slope, $\alpha$, is consistent with that of the
local cluster H$\alpha$ LFs, the characteristic luminosity, $L^\star$,
is about 6 times (or $\approx 2$ mag) brighter.
This strong evolution implies that Abell 521 contains more active
star-forming galaxies than the local clusters, being consistent
with the observed Butcher-Oemler effect.
However, the bright $L^\star$ of Abell 521 may be,
at least in part, due to the dynamical condition of this cluster.

\end{abstract}


\section{INTRODUCTION}

Recent observations of clusters of galaxies at intermediate-redshift
($0.1 \lesssim z \lesssim 0.5$) demonstrated that environmental
processes play an important role on both the star-formation activity and
the morphology of galaxies (e.g., Dressler et al. 1997; Balogh et
al. 1998; Poggianti et al. 1999; Moss \& Whittle 2000).
The galaxy evolution in such clusters may be explained as the
result of the increased activity of the field galaxies at more distant
redshift (Lilly et al. 1996), modulated by changing the infall rate
onto the clusters (Bower 1991).
However, little information on the global star formation history in such
intermediate-redshift clusters has been obtained even to date.

In many previous spectroscopic studies such as the Canadian Network for
Observational Cosmology (CNOC) and MORPHS surveys (e.g., Abraham et
al. 1996; Poggianti et al. 1999), the [O {\sc ii}]$\lambda 3727$
emission line has often been used to investigate the star formation
activity in cluster galaxies.
However, the H$\alpha$ emission line provides a more reliable indicator
of star formation because it is less sensitive both to dust
extinction and to metallicty effects (Kennicutt 1998).
Recently, an H$\alpha$ survey has been made for nearby clusters of 
galaxies Abell 1367 and Coma (Iglesias-P\'{a}ramo et al. 2002).
These data are utilized to derive the H$\alpha$ luminosity function (LF)
of the cluster member galaxies.
Optical spectroscopic surveys are also useful in studying H$\alpha$
emission-line properties of cluster member galaxies (Balogh et al. 2002;
Couch et al. 2001).
However, such spectroscopic sample are selected by using optical
broad-band color properties, being independent from their H$\alpha$
emission-line properties.
Therefore, these data sets may miss a part of H$\alpha$-emitting galaxies
that are too faint to be selected in a continuum-magnitude-limited sample.
In order to determine the shape of the H$\alpha$ LF unambiguously, it is
required to carry out a deeper H$\alpha$ imaging survey.

Motivated by this, we performed our deep H$\alpha$ imaging survey
of a cluster of galaxies, Abell 521 at $z = 0.25$ which is 
a relatively rich (richness class 1) cluster; note that its
Bautz-Morgan Type is III (Abell, Corwin, \& Olowin 1989).
Recent studies of this cluster showed that the projected galaxy density
distribution in the central region has a very anisotropic morphology,
as it exhibits two high-density filaments crossing in an X-shaped
structure at the barycenter of the cluster  (Arnaud et al. 2000; Maurogordato
et al. 2000; Ferrari et al. 2003). The observed
high velocity dispersion of the cluster ($\approx$ 1325 km s$^{-1}$) may be
attributed to the presence of those substructures.
Since this velocity dispersion is also high compared to that expected from
the temperature of the X-ray gas ($kT = 6.3$ keV, Arnaud et al. 2000), 
it is suggested that this cluster is undergoing strong dynamical evolution 
driven by mergers among substructures.
In this paper, we present observational properties of
H$\alpha$-emitting galaxies in Abell 521 and then derive the H$\alpha$
LF for the first time.
Throughout this paper, magnitudes are given in the AB system (Fukugita,
Shimasaku, \& Ichikawa 1995).
We adopt $H_0 = 50$ km s$^{-1}$ Mpc$^{-1}$, $q_0 = 0.5$, and $\Lambda =
0$ cosmology, for comparison with previous results in the literature.

\section{OBSERVATIONS AND DATA REDUCTION}

The imaging data of Abell 521 were obtained with Suprime-Cam (Miyazaki et
al. 2002) on the 8.2 m Subaru Telescope atop Mauna Kea during two runs
(2001 October and 2002 February).
We used the narrow-band filter, $NB816$ ($\lambda_0
= 8150$\AA\ and\ $\Delta \lambda = 120$\AA); the central wavelength
corresponds to a redshift of $\approx 0.24$ for H$\alpha$ emission.
We also used broad-band filters, $B$, $V$, $R_{\rm C}$, $i^\prime$, and
$z^\prime$.
A journal of our observations is summarized in Table \ref{obs}.

We used IRAF and the mosaic-CCD data reduction software, NEKOSOFT
(Yagi et al. 2002), to reduce and combine the individual CCD data.
Photometric and spectrophotometric standard stars used for the flux
calibration are as follows; SA98 (Landolt 1992) for the $B$ data, SA95
and SA98 (Landolt 1992) for the $V$ data, SA95, SA113, and PG0231 (Landolt
1992) for the $R_{\rm C}$ data, GD50 (Oke 1990) for the $i^\prime$ data,
and GD50, GD108, and PG1034+001 (Stone 1996) for the $NB816$ data.
Since the zero point of Landolt (1992) is based on Vega,
we convert the $B$, $V$, and $R_{\rm C}$ magnitudes to the AB system
assuming monotonic colors of $B({\rm Vega}) - B ({\rm AB})=0.140$,
$V({\rm Vega}) - V({\rm AB}) = 0.019$, and $R_{\rm C} ({\rm Vega}) -
R_{\rm C} ({\rm AB}) = -0.169$ (Fukugita et al. 1995).
For $i^\prime$ and $NB816$, we multiplied the atmospheric,
instrumental, and filter transmittance, mirror reflectivity
and CCD quantum efficiency with the SED of standard stars to
calculated absolute flux.
The $z^\prime$ data were calibrated by using the magnitude of the quasar
SDSSp J104433.04--012502.2 (Fan et al. 2000), which were obtained in the
same observing night (see Fujita et al. 2003).
Since any quasar is a potentially variable object, the photometric
calibration of the $z^\prime$ data may be more unreliable than those of
the other band data. 
In order to estimate the uncertainty, we compared the colors of stars
in our area with Galactic stars given by Gunn \&
Stryker\footnote{Bruzual-Persson-Gunn-Stryker (BPGS) Library
(ftp://ftp.stsci.edu/cdbs/cdbs2/grid/bpgs)} and applied the correction to
our $z^\prime$ data.
The combined images for the individual bands were aligned and smoothed
with Gaussian kernels to match the PSF ($= 1\arcsec.1$). The final
images cover an area of 32 $\times$ 20 arcmin$^2$, which corresponds
to 9.4 $\times$ 5.8 Mpc$^2$ at $z = 0.25$.

Source detection and photometry were performed by using SExtractor version
2.2.2 (Bertin \& Arnouts 1996).
The $NB816$ image was used as a reference image for detection, and
photometry was performed for each band image. 
We used a criterion that a source has at least a 13-pixel
connection at $2\sigma$ level.
For each object, we measured the magnitude within a fixed aperture of
a $3\arcsec$ ($\approx$ 15 kpc at $z = 0.25$) diameter for colors and
the magnitudes in all 6 band images.
However, we used an adaptive aperture of diameter $2.5 r_{\rm K}$ for
measurement of total magnitudes, where $r_{\rm K}$ is the Kron radius
(Kron 1980).
Our $3 \sigma_{\rm rms}$ within a $3 \arcsec \phi$ aperture in each band
is given in Table 1.

In order to correct for the Galactic extinction, 
we calculated the extinction at each band (Cardelli et al. 1989),
assuming that $A_B = 0.324$ (Schlegel et al. 1998,
NED\footnote{NASA/IPAC Extragalactic Database
http://nedwww.ipac.caltech.edu/}).
The extinction at each band, $A_\lambda$, is given in Table 1.

\section{SELECTION OF H$\alpha$ EMITTERS}

\begin{figure}
\plotone{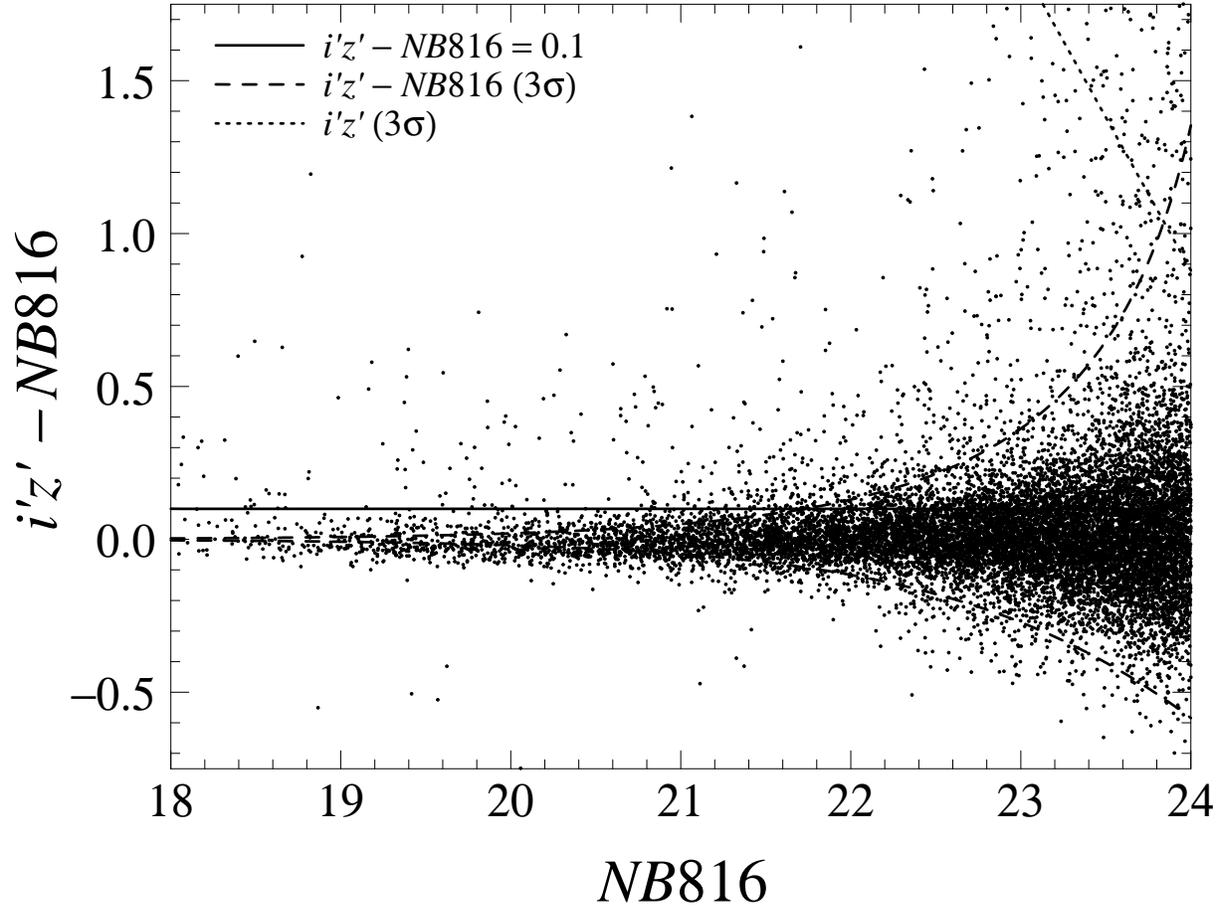}
\caption{The color-magnitude diagram of $i^\prime z^\prime -NB816$
vs. $NB816$.
The horizontal solid line corresponds to $i^\prime z^\prime - NB816 = 0.1$.
The dashed line shows the distribution of $3 \sigma$ error, and the dotted line
shows the limiting $i^\prime z^\prime$ magnitude.
\label{izNB-NB}}
\end{figure}

We defined the off-band continuum flux of objects as $i^\prime z^\prime
\equiv 0.6 i^\prime + 0.4 z^\prime$, which was computed from a linear
interpolation between the effective frequencies of the $i^\prime$ and
$z^\prime$ bands.
Taking the scatter in the $i^\prime z^\prime - NB816$ color into
account, emission-line objects were selected with the following
criteria,
\begin{eqnarray}
i^\prime z^\prime - NB816 &>& {\rm max}\ (0.1,\ 3 \sigma\ {\rm error\ of}\ 
i^\prime z^\prime - NB816), \\
i^\prime z^\prime &<& 24.9.
\end{eqnarray}
The color of $i^\prime z^\prime - NB816 = 0.1$ corresponds to an
equivalent width of $\approx 12$\AA\ in the rest frame at $z = 0.247$.
The magnitude limit corresponds to the $3 \sigma_{\rm rms}$ of the
$i^\prime z^\prime$ band.
These criteria are shown by the solid, dashed, and dotted lines in Figure \ref{izNB-NB}.
Then we found 324 objects that satisfy the above criteria;
note that they are brighter than the limiting magnitude at each band.

\begin{figure}
\plotone{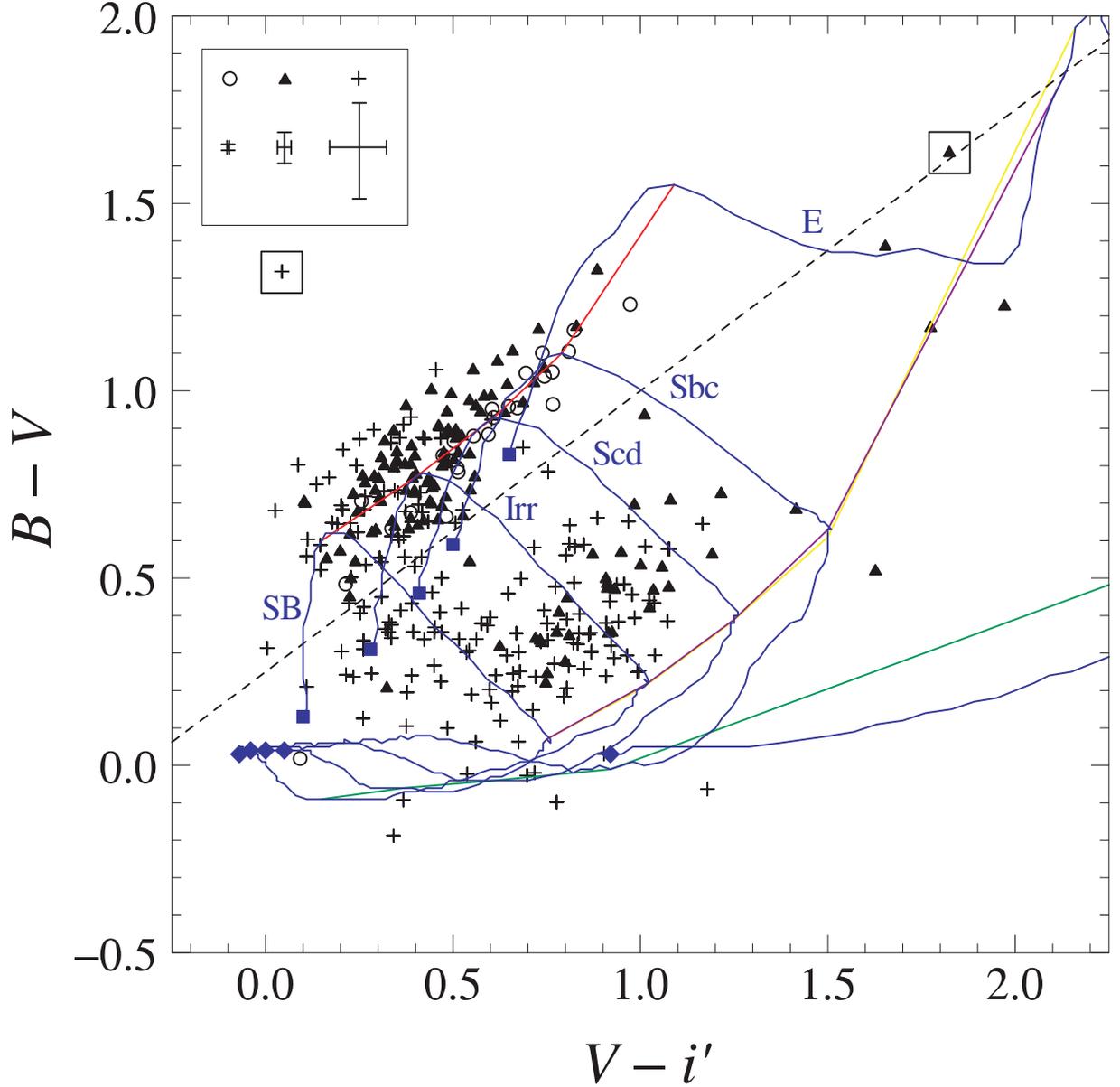}
\caption{The two-color diagram of $B - V$ vs. $V - i^\prime$ for all the
emitters.
The objects with $i^\prime z^\prime < 20$ are shown by open circles,
those with $20 < i^\prime z^\prime < 22$ are shown by filled triangles, and
 those with $i^\prime z^\prime > 22$ are shown by crosses.
The error bars in the left corners of the plots indicate the mean
 uncertainties on colors.
Colors of model galaxies (for SB, Irr, Scd, Sbc, and E) from $z = 0$
(filled squares) to $z = 2$ (filled diamonds) are shown by blue lines.
Colors  of galaxies at $z = 0.24$, 0.64, 0.68, and 1.18 (for H$\alpha$,
[O {\sc iii}], H$\beta$, and [O {\sc ii}] emitters, respectively) are 
shown with red, purple, yellow, and green lines, respectively.
We select the objects above the dashed line as H$\alpha$ emitters except
for the two open squares.
\label{BV-Vi}}
\end{figure}

A narrow band imaging can potentially detect
galaxies with different emission lines at different redshifts.
Emission lines detected in our narrow-band imaging survey
are, for example, H$\alpha$ ($z \approx 0.24$), [O {\sc iii}]$\lambda \lambda 4959, 
5007$ ($z \approx 0.63$), H$\beta$ ($z \approx 0.68$), and 
[O {\sc ii}]$\lambda 3727$ ($z \approx 1.18$).
In order to distinguish H$\alpha$ emitters at $z \approx 0.24$ from other
emission-line objects, we investigate their broad-band color properties. 
In Figure \ref{BV-Vi}, we show the diagram between $B-V$ and $V-i^\prime$
for the 324 emission-line objects.
We also show the model galaxy colors that estimated by using the population
synthesis model GISSEL96 (Bruzual \& Charlot 1993).
The star formation rate of model galaxies is proportional to $\exp
(-t / \tau)$, where we use $\tau = 1$ Gyr.
Changing the age of model galaxies, we generate the SEDs of five types
of galaxies [$t =$ 1, 2, 3, 4, and 8 Gyr for SB (young starburst), Irr,
Scd, Sbc, and E, respectively].
Then we find that H$\alpha$ emitters at $z \approx 0.24$ can be
selected by adopting the following color criterion, 
\begin{equation}
B - V > 0.75 (V - i^\prime) + 0.25.
\end{equation}
As shown in Figure 2, this criterion appears to isolate H$\alpha$
emitters from other emitters at different redshifts.
We then obtain a sample of 165 H$\alpha$ emitters.

In order to examine whether or not this color criterion works well more
quantitatively, we investigate color properties of the spectroscopically
confirmed member galaxies in Abell 521.
The previous spectroscopic surveys identified 125 member galaxies in
this cluster (Ferrari et al. 2003; Maurogordato et al. 2000).
These members are identified in our catalog. 
We found that 123 objects satisfy the color criterion based on our photometry.
The remaining two objects were rejected in our selection procedure
because they have neighboring galaxies, causing large photometric errors.
As a further check, we examined the spatial distribution of the galaxies
which are selected by the above color criterion.
In Figure \ref{density}, the excess of  surface number density by the
cluster members is clearly seen in the area within the radial distance
of 2 Mpc from the center
of Abell 521;  note that the center is taken at the position of the main
X-ray cluster (Arnaud et al. 2000).

In order to investigate the detection completeness of H$\alpha$ emitters,
we performed a detection test on simulated artificial images.
We prepared model galaxies with broad-band colors of late type galaxies
(SB, Irr, Scd, and Sbc) at $z = 0.24$ generated by GISSEL96 model.
Since the equivalent width of H$\alpha$ emission is generally different form
galaxy to galaxy, we put the excess of $NB816$ to $i^\prime z^\prime$
continuum as a parameter.
We studied a color range of $0.1 < i^\prime z^\prime - NB816 < 1.1$ with
five bins of $\Delta (i^\prime z^\prime - NB816) = 0.2$.
Since the total magnitude of $NB816$ is another parameter, we studied a
magnitude range of $18.0 < NB816 < 24.0$ with twelve bins of $\Delta(NB816) = 0.5$.
We prepared 500 model galaxies in each set of $NB816$ and $i^\prime
z^\prime - NB816$.
The light distribution of model galaxies is generated as the exponential
law by IRAF ARTDATA.
Their sky positions, half-light radius (1 to 7 kpc), and ellipticities
(0.3 to 1.0) were randomly set and put into the CCD data together with
Poisson noises.
After smoothing model-galaxy images to match to the seeing size, we tried
to detect them using SExtractor and to select as H$\alpha$ emitters with
the same procedure as that used for the observed data.
The detectability of the model emitters is shown in Figure \ref{comp}
as a function of $NB816$ magnitude.
We found that the detectability is higher than 80\% for objects with
$NB816 < 22.0$, but lower than 30\% for objects with $23.0 < NB816 < 24.0$.

Next, we examined the contamination by emission-line objects at
other redshifts (e.g., [O {\sc iii}]$\lambda \lambda 4959,5007$ at $z
\approx 0.63$) in the same manner as that for the detection completeness
estimate.
We found the contamination to be only less than 5\% even for the faint
objects with $23.0 < NB816 < 24.0$.
In conclusion, our color criterion allows us to select reliable
H$\alpha$ emitters at $z \approx 0.24$ for sources with $NB816 < 23.0$.

\section{H$\alpha$ LUMINOSITY FUNCTION}

\subsection{Emission-Line Equivalent Width}

At first we derive the emission-line equivalent width of the individual
H$\alpha$ emitters.
Since the pass band of our narrow-band filter
is too wide to separate [N {\sc ii}]$\lambda \lambda 6548,
6584$ from H$\alpha$, the equivalent with derived here is 
that contributed from the three emission lines.
We adopt the same method as that used by Pascual et al. (2001) to
calculate H$\alpha$ + [N {\sc ii}]$\lambda \lambda 6548, 6584$
(hereafter H$\alpha$ + [N {\sc ii}]) equivalent width.
The flux density in each filter can be expressed as the sum of the line
flux and the continuum density (the line is covered by both filters),
\begin{eqnarray}
f_{NB} &=& f_{\rm C} + (F_{\rm L} / \Delta NB), \\
f_{i^\prime z^\prime} &=& f_{\rm C} + 0.6 (F_{\rm L} / \Delta i^\prime), 
\end{eqnarray}
where $f_{\rm C}$ is the continuum flux density, $F_{\rm L}$ the line flux,
$\Delta NB$ and $\Delta i^\prime$ the $NB816$ and $i^\prime$ band filter
effective widths, respectively,  and $f_{NB}$ and $f_{i^\prime
z^\prime}$ the flux density in each filter.
Then the equivalent width can be expressed as follows;
\begin{equation}
EW_{\rm obs} ({\rm H} \alpha + [\textrm{\ion{N}{2}}]) 
= \frac{F_{\rm L}}{f_{\rm C}}
= \Delta NB \left[ \frac{f_{NB} - f_{i^\prime z^\prime}}{f_{i^\prime z^\prime} -
0.6  f_{NB} (\Delta NB / \Delta i^\prime) } \right],
\end{equation}
where $\Delta i^\prime = 1535$\AA\ for the $i^\prime$ band and $\Delta
NB = 120$\AA\ for the $NB816$ band.

\subsection{H$\alpha$ Luminosity}

In order to obtain the H$\alpha$ luminosity, we have to correct for
the flux contribution from the [N {\sc ii}] doublet.
We also have to apply correction for internal extinction. 
For the two corrections, we adopt the flux ratio of
H$\alpha$/[N {\sc ii}] = 2.3 (Kennicutt 1992) and
$A_{{\rm H}\alpha} = 1$ mag as a typical internal extinction at $z \sim 0.2$
(Tresse \& Maddox 1998).
Applying these corrections, we estimate the H$\alpha$ luminosity, $L ({\rm
H}\alpha)$, from the total magnitude for each objects.
We then obtain the star formation rate, $SFR$, using 
the following relation (Kennicutt 1998), 
\begin{equation}
SFR\ (M_\odot\ {\rm yr}^{-1}) = 7.9 \times 10^{-42} L ({\rm H} \alpha)\ ({\rm erg\ s}^{-1}).
\end{equation}

The net H$\alpha$ flux depends on the source redshift because of the
fixed transmittance of our $NB816$ filter.
In order to correct for this effect, we adopt the following assumptions.
(1) The velocity (redshift) distribution of the cluster galaxies is a
Gaussian probability density function $P (z)$ around a mean velocity of
74019 km s$^{-1}$ with a dispersion of 1325 km s$^{-1}$ (Ferrari et al. 2003).
(2) The equivalent width distribution of the H$\alpha$
emitters in the rest frame is expressed as the following exponential
functional form; $Q(EW) = A \exp ( - EW / B )$, where $A$ and $B$ are
free parameters.
Then, the rest frame equivalent width probability of the
H$\alpha$ emitters at redshift $z$ can be expressed as $P(z) \times Q(EW)$.
The best-fit parameters are obtained so as to fit
the distribution of $i^\prime z^\prime - NB816$.

We apply the correction for the incomplete velocity
coverage of the $NB816$ filter.
Since the number of the H$\alpha$ emitters excluded from the filter can be
calculated from the value of the equivalent width, 
we correct the equivalent width distribution function that derived above.
Therefore, using the equivalent width distribution function,
we obtain the corrected H$\alpha$ luminosity distribution.

\subsection{Statistical Weights}

Since the excess of surface galaxy density by the cluster member is
clearly seen in the area within 2 Mpc radius (see Figure 3), it seems
safe to use the H$\alpha$ emitters located in the region;
note that there are 53 H$\alpha$ emitters in this region.
Although the $NB816$ filter corresponds to the redshifted H$\alpha$
emission line at $z = 0.242 \pm 0.009$, the mean redshift of the cluster
is $z = 0.247$.
Therefore, our H$\alpha$ emitter sample may suffer 
contamination from field H$\alpha$ emitters,
in particular from field galaxies in the foreground.
In order to correct for this effect, we estimate the field
contamination by the following method.

In order to obtain a sample of field galaxies, we use an area 
of 3 Mpc outside from the cluster center, taking account that the virial
radius is estimated to be 2.74 Mpc (Girardi \& Mezzetti 2001).
However, since a subclump is seen to the North-East
at about 4.5  Mpc from the cluster center, we exclude this area
from our analysis. 
The remaining sky area is used to estimate the surface number density of
field galaxies as a function of $NB816$ magnitude.
Since Abell 521 shows elongated galaxy distribution (Ferrari et
al. 2003), we use the local surface number density, which is calculated
from 10 nearest neighbors, rather than the radial surface number density
distribution.
In this way, all H$\alpha$ emitters in the cluster region are weighted by the
luminosity-dependent factor, $w_{\rm l}$ and the density-dependent factor, 
$w_{\rm d}$, after renormalizing $w_{\rm d}$ so that the sum of the 
total galaxy weights, $w_{\rm l} \times w_{\rm d}$ is equal to the sum 
of $w_{\rm l}$.

\subsection{H$\alpha$ Luminosity Function}

We derive the H$\alpha$ luminosity function (LF) for the cluster region.
In order to normalize the H$\alpha$ LF to a proper volume, we use the
sphere volume within 2 Mpc radius.
The H$\alpha$ LF for our sample is well fitted by the Schechter (1976)
function;
\begin{equation}
\phi (L) {\rm d}L = \phi^\star (L/L^\star)^\alpha \exp (-L/L^\star) 
{\rm d}(L/L^\star).
\end{equation}
The size of the bins is taken to be $\Delta \log L ({\rm H} \alpha) =
0.5$.
The number counts in the individual bins are given in Table \ref{count},
and  Figure \ref{HAE_LF1} shows the H$\alpha$ LF for Abell 521 and its
Schechter fit.
After correcting for the detection completeness (see section 3), we
obtain the best-fit parameters for the cluster membership within 2 Mpc
radius are
\begin{eqnarray}
\alpha &=& - 0.75 \pm 0.23, \nonumber \\
\phi^\star &=& 10^{-0.25 \pm 0.20} \ {\rm Mpc}^{-3}, \nonumber \\
L^\star &=& 10^{42.03 \pm 0.17}\ {\rm erg\ s}^{-1}. \nonumber
\end{eqnarray}
The inset panel in Figure \ref{HAE_LF1} shows the $\chi^2$ error
contours corresponding to the 1 and 2$\sigma$ levels.
It can be seen that $\alpha$ and $L^\star$ values are highly correlated.

\section{DISCUSSION}

In Figure \ref{HAE_LF1}, we also show the H$\alpha$ LFs of field
galaxies in the local Universe (Gallego et al. 1995), at $z \sim 0.2$
(Tresse \& Maddox 1998), at $z \approx 0.24$ (Fujita et al. 2003), and
those of the nearby clusters of galaxies (Iglesias-P\'{a}ramo et al
2002).
As for the nearby clusters of galaxies, Iglesias-P\'{a}ramo et al. (2002)
derived the H$\alpha$ LFs for Abell 1367 and Coma (richness class 2).
The physical areas analysed by them are nearly the same as that of Abell
521.
Although they applied a slightly different extinction correction; $A_{{\rm
H} \alpha} = 1.1$ mag for type Scd or earlier and $A_{{\rm H} \alpha} =
0.6$ for type Sd or later (Boselli et al. 2001), they obtained
$\alpha \approx -0.7$ and $L^\star \approx 10^{41.25}$ erg s$^{-1}$
for these clusters.
The inset panel in this figure shows the best-fit parameters of Abell
521 and the others.
There is the significant difference in H$\alpha$ LFs between the
clusters and the fields.
It is found that the faint end slope, $\alpha$,
of the H$\alpha$ LF of Abell 521 is flatter than that of the field H$\alpha$
LFs at the same redshift.
Therefore, although the sample analysed  here is not so large, it appears that
the flatter faint end slope of the H$\alpha$ LF is commonly
seen in the clusters of galaxies between $z \simeq 0$ and $z \simeq 0.2$.

The most important finding in this study is that the characteristic
luminosity, $L^\star$ of Abell 521 is $\approx$ 6 times (or $\approx$
2 mag) brighter than those of the local cluster H$\alpha$ LFs, and nearly
the same as those of the field H$\alpha$ LFs.
The evolution of the field H$\alpha$ LF from $z \simeq 0.2$
to $z \simeq 0$ appears closely related to the $\phi^\star$ parameter
rather than to the $L^\star$ and $\alpha$ parameters.
However, as for the cluster H$\alpha$ LFs, the $L^\star$ parameter
appears to evolve strongly from $z \simeq 0.2$ to $z \simeq 0$.
Furthermore, we investigate the following galaxy clusters for which
spectroscopic H$\alpha$ emitter surveys were already
made; Abell 1689 at $z = 0.18$ (Balogh et al. 2002) and AC 114 at $z =
0.31$ (Couch et al. 2001).
Since their observations are based on spectroscopy,
it is difficult to compare with our H$\alpha$ LF straightforwardly.
Therefore, we compare the H$\alpha$ LF of Abell 521
with the other clusters, only using the H$\alpha$ luminous galaxies
($L^\star \gtrsim 10^{41.5}$ erg s$^{-1}$).
It should be noted that the same dust extinction correction (i.e.,
$A_{{\rm H} \alpha} = 1$ mag) is applied for these clusters, and the
H$\alpha$ luminosity from spectroscopy may be underestimated due to
aperture bias.

Figure \ref{HAE_LF2} shows the surface density of H$\alpha$ luminosity
distribution of each cluster; note that the surface density is used
instead of the volume density which is used in Figure 5, to compare
simply with previous results.
Although there are some galaxies whose $SFR$s
exceed $SFR\ \sim 4 M_\odot$ yr$^{-1}$ in Abell 521 and Abell 1689, no
such galaxies are found in Abell 1367 and Coma.
This $SFR$ is about the same level as that observed in the Milky Way
(Rana 1991).
This result implies that Abell 521 and Abell 1689 contain more active
star-forming galaxies than the nearby clusters.
However, AC 114 has nearly the same property in the H$\alpha$ LF as
those of the nearby clusters in spite of its higher redshift.

This evolution may simply reflect the Butcher-Oemler (BO) effect, which
is the increase in the fraction of blue galaxies, presumably
star-forming galaxies, in clusters with increasing redshift (Butcher
\& Oemler 1984).
In fact, the fraction of blue galaxies in our H$\alpha$ emitters by the
definition of Butcher \& Oemler (1984) is $\sim$ 85\%.
The blue fraction, $f_{\rm B}$, in Abell 1689 is comparable to that
of Abell 521 ($f_{\rm B} \sim 0.17$, see for details Appendix) and much
higher than those in the nearby rich clusters such as Coma ($f_{\rm B} =
0.03$).
However, AC 114 and Abell 1367 also have a large fraction of blue galaxies
($f_{\rm B} \sim 0.20$).
Therefore, the star formation activity in clusters of galaxies may not
be understood solely by the BO effect.

These results suggest that some physical processes work
in their star formation activity, independent of the blue fraction of
each cluster.
We investigated what parameter of cluster correlates with
the existence of high $SFR$ galaxies.
One possible parameter is the dynamical status, as pointed out by Balogh
et al. (2002).
Although Abell 1689 appears to show a round shape in
X-ray, the observed radial velocity distributions of the cluster shows
substructures.
On the other hand, AC 114 appears to be relaxed dynamically, although the
cluster member galaxies shows an elongated morphology (Balogh et
al. 2002).
Therefore, it seems important to investigate the radial velocity
distributions for the H$\alpha$ emitters
of each cluster (Figure \ref{vr_dist}).
In this figures, we use the relative radial velociy, $\Delta v_{\rm r}$,
to the systemic velocity, $v_{\rm r,0}$, of each cluster, and $v_{\rm
r,0}$ is shown in the left corners of each figure.
For Abell 521, Abell 1367, and Coma, their redshift are unknown from
their imaging survey, since the H$\alpha$ emitters are detected using
the narrow-band filters.
Therefore, their distributions are shown using available
spectroscopic data in the literature (Ferrari et al. 2003;
Iglesias-P\'{a}ramo et al. 2002; Cortese et al. 2003).
The vertical dashed lines in this figures show the velocity coverage
corresponding to the $FWHM$ of the narrow-band filters.
For Abell 521, in particular, since some H$\alpha$ emitters with
$\Delta v_{\rm r} > 0$ km s$^{-1}$ may be excluded from the filter
transmittance, we also show the velocity distribution of galaxies
identified with other emission lines ([\ion{O}{2}] or H$\beta$) taken
from Ferrari et al. (2003).
It is found that this radial velocity distribution is consistent with
that of the H$\alpha$ emitters with $\Delta v_{\rm r} < 0$ km s$^{-1}$.
Therefore, we expect that it also traces that of the H$\alpha$
emitters with $\Delta v_{\rm r} > 0$ km s$^{-1}$.
In this figure, the radial velocity distribution for the H$\alpha$ emitters in
Abell 521 is inconsistent with a Gaussian distribution which we assumed
in section 4.1.
However, since the spectroscopic sample is rather small, it is dangerous
to apply this distribution to our sample.
Even if this distribution is valid, it does not change the result that
$L^\star$ of Abell 521 is much brighter than those of the nearby clusters.

In Figure \ref{vr_dist}, the radial velocity distributions in Abell 1367
and Coma tend to be concentrated around $\Delta v_{\rm r} \approx 0$ km
s$^{-1}$.
Therefore, it is suggested that these galaxies are associated with a
galaxy population in the cluster core, probably a virialized system.
For AC 114, the radial velocity distribution suggests that the most
H$\alpha$ emitters are associated with the structure centered at $\Delta
v_{\rm r} \approx +2000$ km s$^{-1}$.
Even if they comprise a substructure infalling on to the cluster center, the
star-forming activity seems to have already been ceasing by the
environment effect (Kodama et al. 2001; G\'{o}mez et al. 2003).
Abell 521 and Abell 1689 show that the radial velocity distributions show
several peaks in a wide velocity range, and there are few H$\alpha$
emitters at the mean cluster redshifts.
And for Abell 521, they are not thought to be field contamination
but be associated with the cluster, since their number density is high
at the center of the cluster (Figure 3).
The H$\alpha$ emitters in the two clusters are likely to be in the
process of accretion of the field population.
If this is the case, we may explain why their star-forming activities are
higher than those of the other clusters.
It is thus suggested that the star formation activity in clusters seems
to be strongly related to the dynamical status of the clusters.

\vspace{0.5cm}

We would like to thank the Suprime-Cam team and the Subaru Telescope
staff for their invaluable help, and T. Hayashino for his technical help.
We also an anonymous referee for his/her useful comments and
suggestions.
This research has made use of the NASA/IPAC Extragalactic Database
(NED) which is operated by the Jet Propulsion Laboratory, California
Institute of Technology, under contract with the National Aeronautics
and Space Administration.
The data reduction is partly done using computer system at
the Astronomical Data Analysis Center of
the National Astronomical Observatory of Japan.
This work was financially supported in part by the Ministry of
Education, Culture, Sports, Science, and Technology (Nos. 1004405,
10304013, and 15340059). TN and MA are JSPS fellows.

\appendix
\section{COMMENTS ON THE BLUE FRACTION}

In order to derive the fraction of blue galaxies, $f_{\rm B}$, in Abell 521,
we followed the procedure of Butcher \& Oemler (1984).
We calculated the fraction of galaxies that are 0.2 mag bluer than the
color-magnitude sequence of early-type galaxies in the rest-frame $B -
V$ color within a magnitude limit, $M_V = -20$, and that are enclosed
within $R_{30}$, the radius which contains 30\% of the cluster
population.

\begin{figure}
\plotone{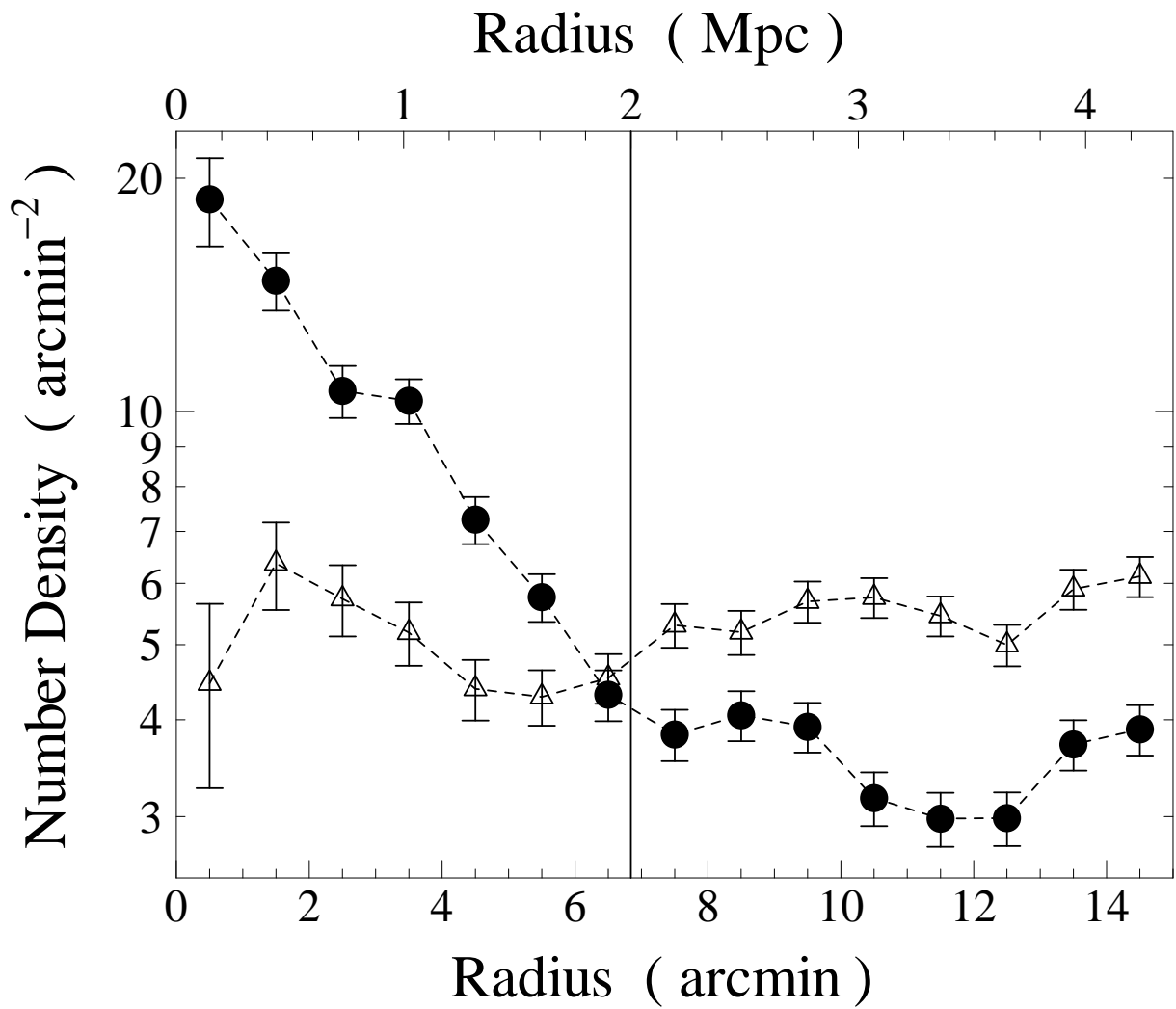}
\caption{Surface number density distribution of the galaxies with
 $i^\prime z^\prime < 23$ as a function of the radius from the center of
 Abell 521.
The objects with $B - V > 0.75 (V - i^\prime) + 0.25$ are shown by the
 filled circles and those with $B - V \leq 0.75 (V - i^\prime) + 0.25$
 are shown by the open triangles.
The vertical solid line shows the radius of 2 Mpc from the cluster center.
\label{density}}
\end{figure}

\begin{figure}
\plotone{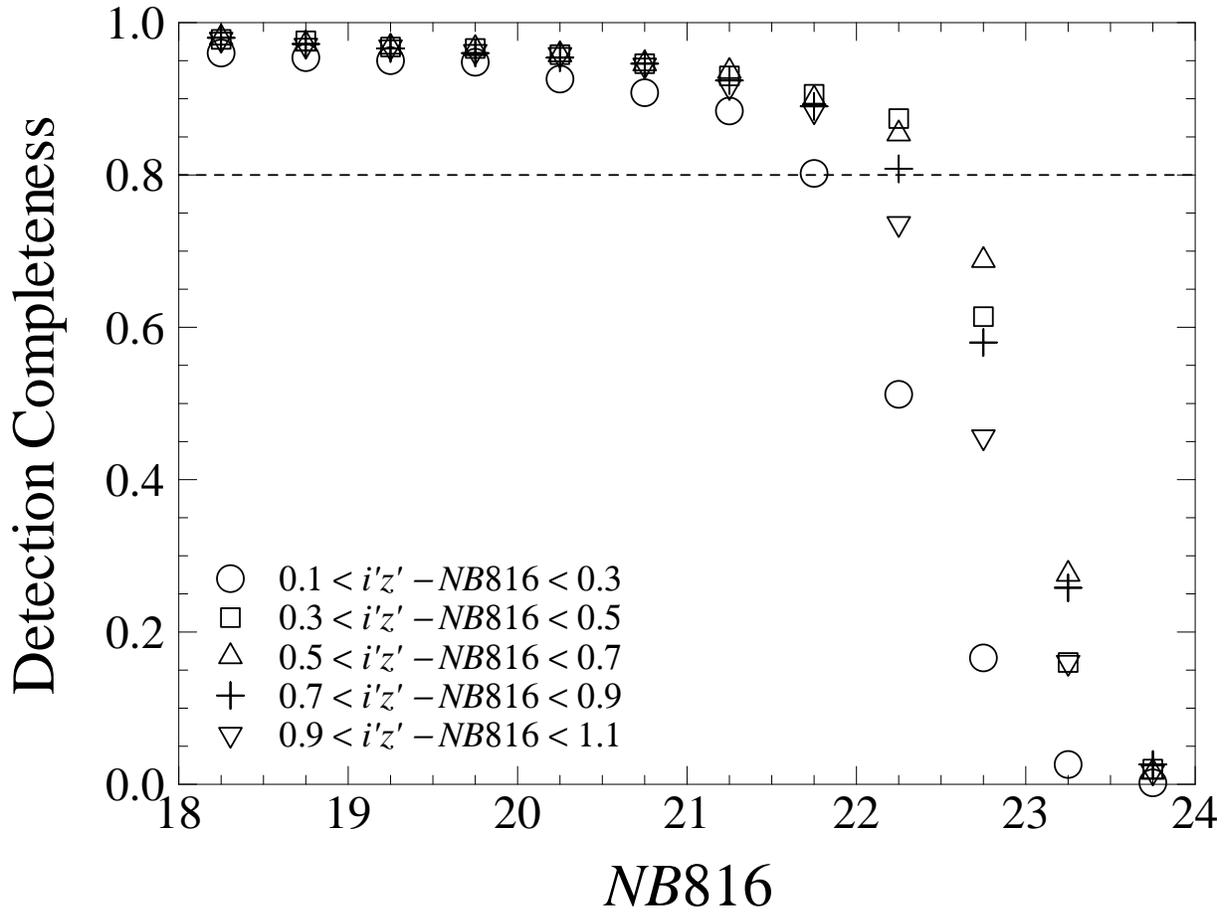}
\caption{Detection completeness of the H$\alpha$ emitters derived from
 the simulation.
The horizontal line shows the detection completeness of 0.80.
\label{comp}}
\end{figure}

\begin{figure}
\plotone{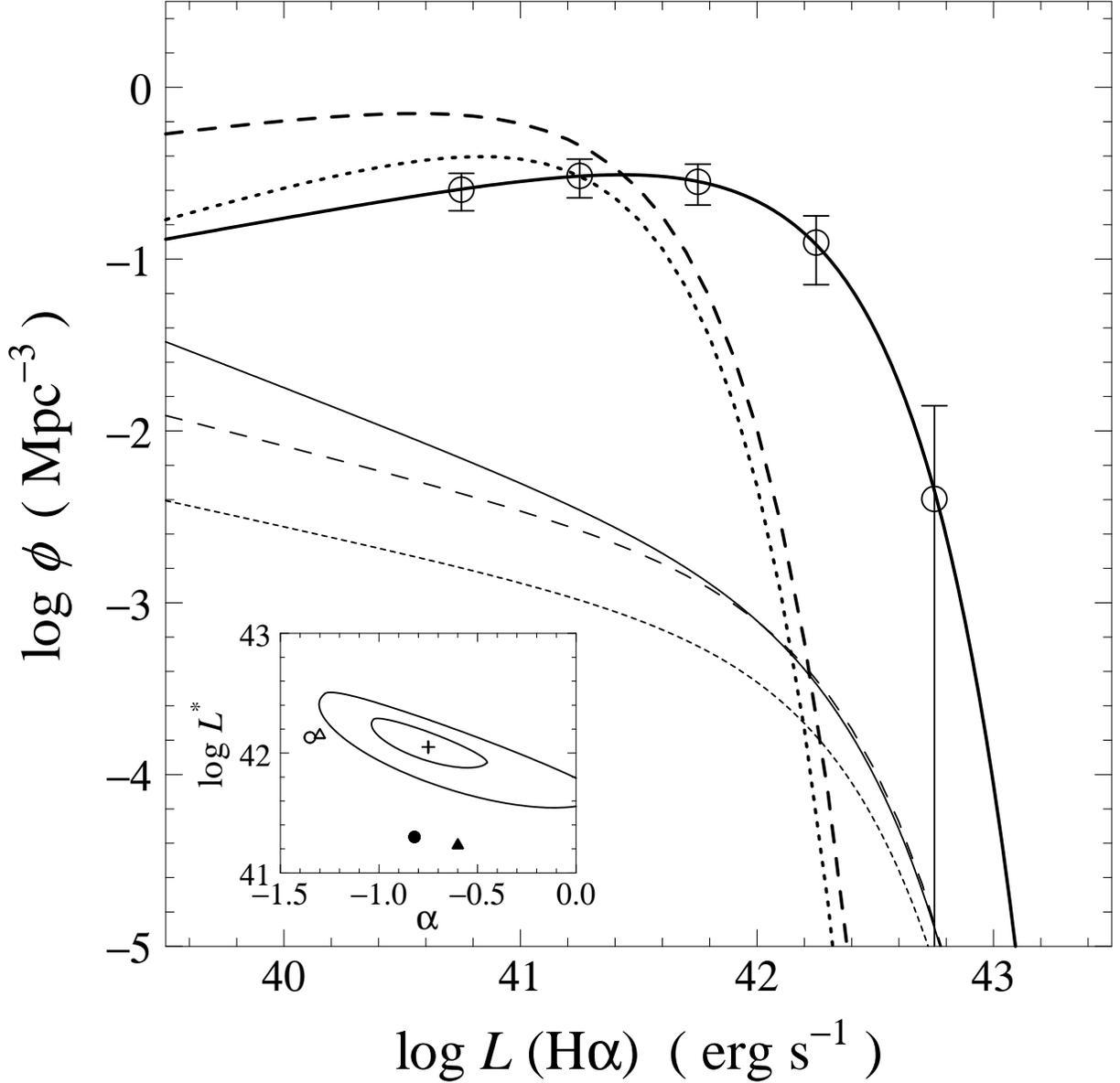}
\caption{The derived H$\alpha$ luminosity function (LF) for Abell 521 is
 shown by the thick solid curve with the data points (open circles).
For reference, the H$\alpha$ LFs of Abell 1367 and Coma
 (Iglesias-P\'{a}ramo et al. 2002) are shown by the thick dashed and
 dotted curves, respectively.
We also show those of field galaxies in the local Universe (Gallego et
 al. 1995, thin dotted line), at $z \sim 0.2$ (Tresse \& Maddox 1998,
 thin dashed line), and at $z \approx 0.24$ (Fujita et al. 2003, thin
 solid line).
Inset panel shows 1 and 2$\sigma$ error contours for the best-fit
 H$\alpha$ LF parameters of Abell 521. 
The best-fit parameters of Abell 1367 and Coma  are shown by the filled
 circle and the filled triangle, and those of field galaxies in the
 local Universe and at $z \sim 0.2$ are shown by the open circle and the
 open triangle, respectively.
Note that those of field galaxies at $z \approx 0.24$ are not shown
 in this figure.
\label{HAE_LF1}}
\end{figure}

\begin{figure}
\plotone{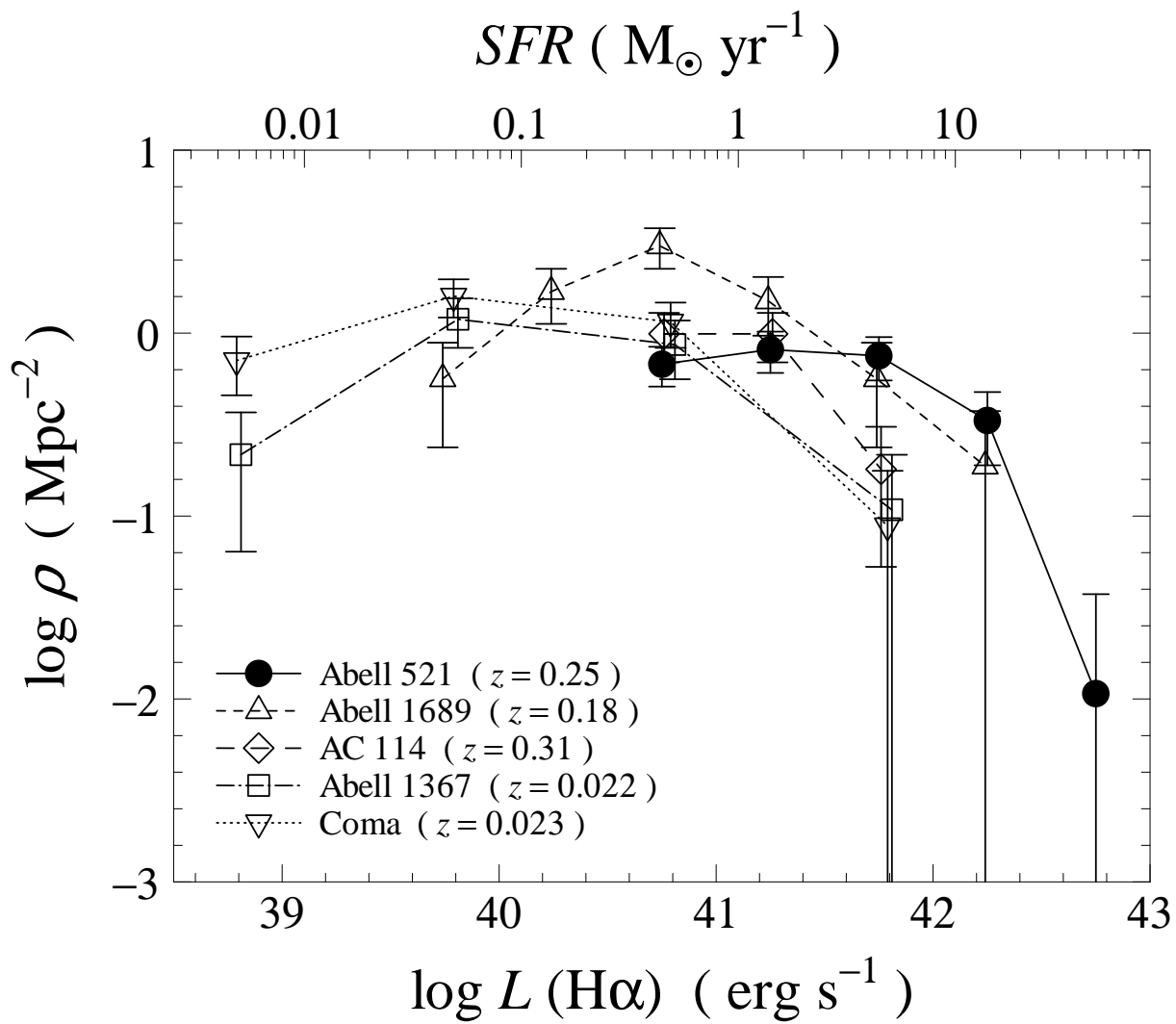}
\caption{The surface density of H$\alpha$ luminosity distribution of the
 cluster galaxies for Abell 521.
We also show those for Abell 1367, Coma (Iglesias-P\'{a}ramo et
 al. 2002), Abell 1689 (Balogh et al. 2002), and AC 114 (Couch et al. 2001).
\label{HAE_LF2}}
\end{figure}

\begin{figure}
\plotone{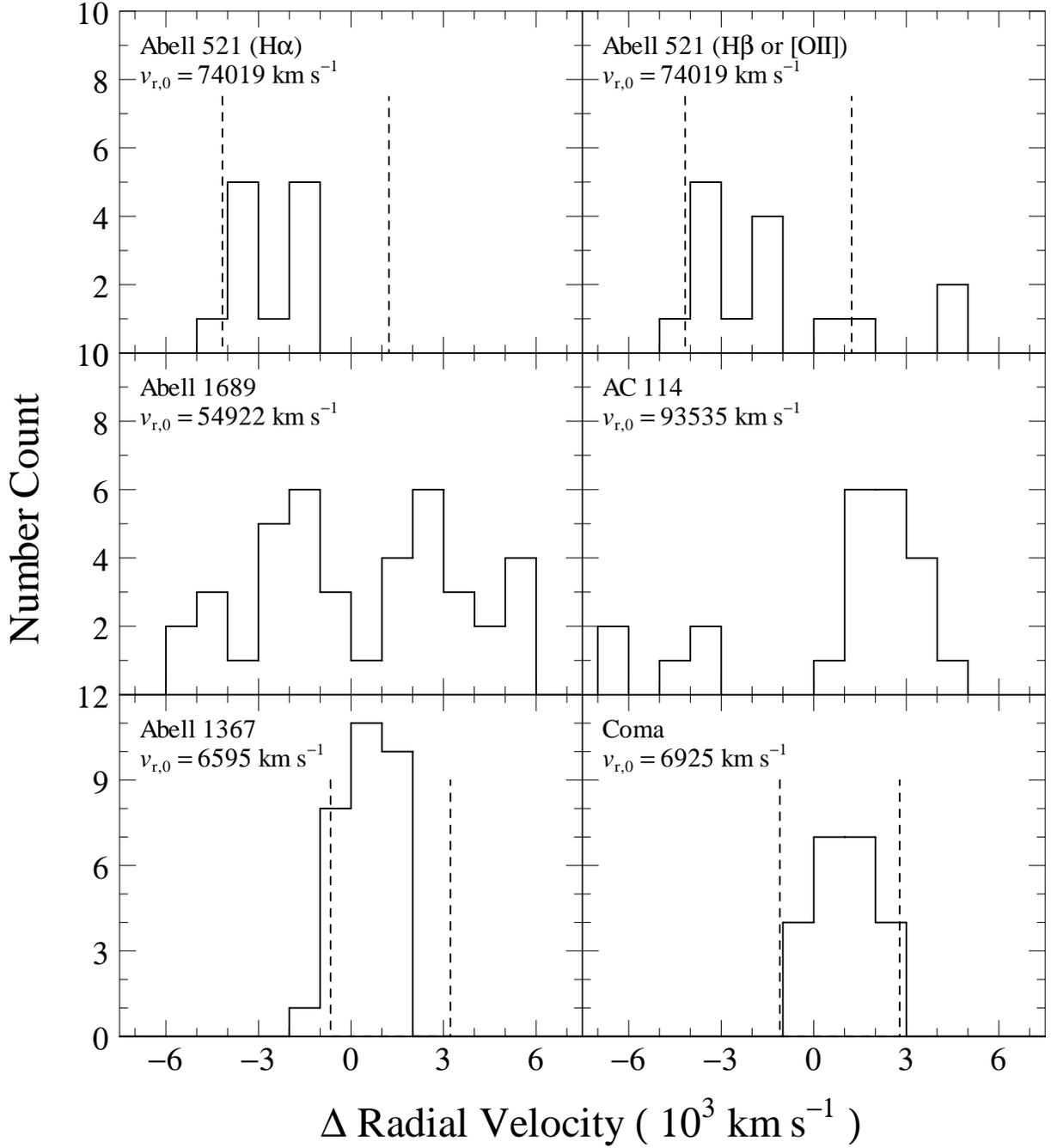}
\caption{Radial velocity distributions for the H$\alpha$ emitters of each
 cluster, with a binning of 1000 km s$^{-1}$.
We use the relative radial velociy to the systemic velocity, $v_{\rm r,0}$, 
of each cluster, and $v_{\rm r,0}$ is shown in the left corners of each figure.
The vertical dashed lines show the velocity coverage corresponding to
 the $FWHM$ of the narrow-band filters for Abell 521, Abell 1367, and Coma.
\label{vr_dist}}
\end{figure}

The blue fraction, $f_{\rm B}$, was calculated by subtracting the number
count expected for the field from that obtained for the cluster sample.
In order to obtain the value of $f_{\rm B}$, we used the $V - R_{\rm C}$
color index, which is very close to $B - V$ rest frame at the cluster
redshift.
The magnitude limit, $M_V = -20$, in the rest frame corresponds to
$R_{\rm C} \simeq 20.8$ at cluster redshift, and a difference of $\Delta
(B - V) = 0.2$ mag is $\Delta (V - R_{\rm C}) \simeq 0.17$ mag.
The value of $R_{30}$ is determined from the surface number density
profile (Figure \ref{density}).
We found $R_{\rm 30} = 3.1$ arcmin ($\approx 0.9$ Mpc), being
very similar to the typical radius of $R_{30} \sim 1$ Mpc for a sample
of rich clusters at $0.2 \lesssim z \lesssim 0.4$ (Kodama \& Bower 2001).
As a result, we obtain $f_{\rm B} = 0.17 \pm 0.03$.

\clearpage

\begin{deluxetable}{ccccc}
\tablecaption{Journal of the observations \label{obs}}
\tablewidth{0pt}
\tablehead{
\colhead{Filter} & \colhead{Obs. Date} & \colhead{Total Exp. (s)} &
 \colhead{$A_\lambda$\tablenotemark{a}} & \colhead{Limiting Mag.\tablenotemark{b}}
}
\startdata
$B$ & 2002 Feb. 15 & 480 & 0.324 & 25.3 \\
$V$ & 2001 Oct. 14, 15 & 2250 & 0.245 & 25.5 \\
$R_{\rm C}$ & 2001 Oct. 14 & 1620 & 0.203 & 25.5 \\ 
$i^\prime$ & 2001 Oct. 14 & 1320 & 0.157 & 25.4 \\ 
$NB816$ & 2002 Feb. 14 & 3600 & 0.139 & 24.6 \\
$z^\prime$ & 2002 Feb. 15 & 840 & 0.113 & 24.2 \\
\enddata
\tablenotetext{a}{Galactic extinction are applied to our imaging data
 (Cardelli et al. 1989; Schlegel et al. 1998).}
\tablenotetext{b}{The limiting AB magnitude (3$\sigma$) within a 
$3 \arcsec \phi$ aperture.}
\end{deluxetable}

\begin{deluxetable}{ccc}
\tablecaption{H$\alpha$ luminosity function \label{count}}
\tablewidth{0pt}
\tablehead{
\colhead{$\log L ({\rm H} \alpha)$ (erg s$^{-1}$)} 
& \colhead{Number of H$\alpha$ emitters} 
& \colhead{$\log \phi$ (Mpc$^{-3}$)}
}
\startdata
40.5 -- 41.0 & 10.62 &  $-0.60$ \\
41.0 -- 41.5 & 12.77 &  $-0.52$ \\
41.5 -- 42.0 & 11.81 &  $-0.55$ \\
42.0 -- 42.5 &  5.23 &  $-0.90$ \\
42.5 -- 43.0 &  0.17 &  $-2.40$ \\
\enddata
\end{deluxetable}

\clearpage

\epsscale{1.0}

\end{document}